\documentclass[aps,pra,twocolumn,showpacs,groupedaddress,10pt]{revtex4-1}
\usepackage{graphicx}

\begin{document}


\title{Electron Evaporation from an Ultracold Plasma in a Uniform Electric Field}


\author{K. A. Twedt}
\email{twedt@umd.edu}
\author{S. L. Rolston}
\affiliation{Joint Quantum Institute and Department of Physics, University of Maryland, College Park, Maryland 20742}


\date{\today}

\begin{abstract}
Electrons in an expanding ultracold plasma are expected to be in quasi-equilibrium, since the collision times are short compared to the plasma lifetime, yet we observe electrons evaporating out as the ion density decreases during expansion.  We observe that a small electric field that shifts the electron cloud with respect to the ions increases the evaporation rate.  We have calculated the spatial distribution of a zero-temperature electron cloud as a function of applied field and ion density, which is assumed to be Gaussian at all times.  This calculation allows us to predict the flux of cold electrons from the plasma at all times, and is in good agreement with our observed electron signal.
\end{abstract}


\maketitle

Ultracold neutral plasmas (UNPs) \cite{killianreview, killian2007} are created by the photoionization of laser-cooled neutral atoms.  They have typical densities of $1 \times 10^9 {\rm cm^{-3}}$ and electron temperatures ranging from 1-1000ÊK.  UNPs are typically unconfined and expand due to electron pressure into the surrounding vacuum with lifetimes of a few hundred microseconds, limited by stray electric fields.  During this expansion, the electron temperature is changing dynamically from the competition of adiabatic and evaporative cooling with various heating processes, such as three-body recombination.  Previous experimental work has focused on measuring the global electron temperature \cite{roberts2004, gupta2007, fletcher2007}, but the thermal distribution of the electrons has also drawn interest \cite{pohl2004, pohl2006, comparat2005, Vrinceanu2008}.  Since the plasma electrons are held in a finite-depth potential well, a true Maxwell-Boltzmann distribution cannot exist, leading to predictions of a Michie-King distribution \cite{comparat2005, Vrinceanu2008}, which has been developed for gravitationally bound systems in astrophysics.  This is in contrast to a truncated Boltzmann distribution that has been proposed for a similar system in quasi-equilibrium, trapped ultracold atoms undergoing forced evaporation for the production of Bose-Einstein condensates \cite{luiten1996}.  Electrons in UNPs may also provide insight into other physical systems, as their behavior shares many similarities with electrons moving in laser-irradiated atomic clusters \cite{rost2009} and stars in globular clusters \cite{comparat2005}.

Probing electron dynamics is difficult in UNPs since the low total number of electrons prevents the use of electric probes inside the plasma.  Instead, electrons are detected by applying a small uniform electric field across the plasma to direct escaping electrons to a charged particle detector.  Electron detection has been used to measure collective oscillations of the plasma \cite{kulin2000, fletcher2006}, Rydberg atom formation \cite{killian2001, fletcher2007}, plasma instabilities \cite{zhang2008b}, and electron temperature \cite{roberts2004, fletcher2007}.  All of these measurements involved applying perturbing fields, from dc pulses to microwaves, and observing the change in the electron emission.  Electron detection has also been a main diagnostic in studies of the spontaneous evolution of Rydberg gases into plasmas \cite{Walz-Flannigan2004, robinson2000} and of a UNP in a strong magnetic trap \cite{choi2008}.  But an explanation of the main feature of the unperturbed electron emission signal is missing from the literature.  

\begin{figure}[t]
\includegraphics[width=3.4in]{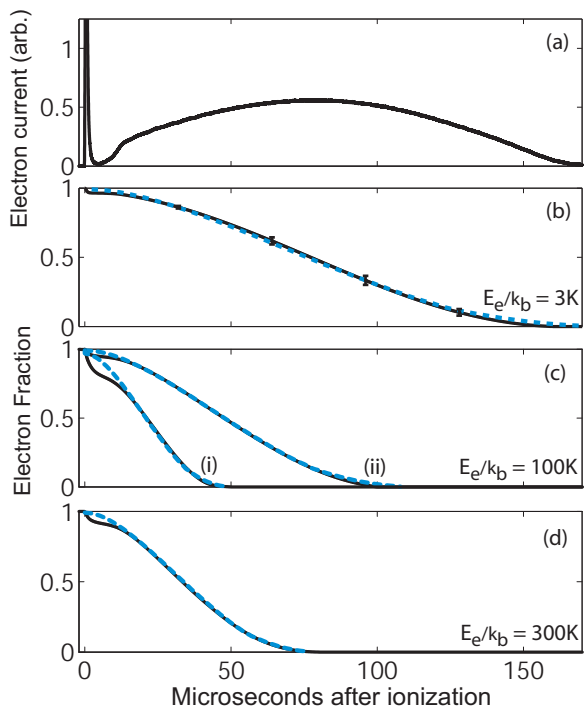}
\vskip -0.2in
\caption{ (a) An example of our canonical electron current signal. (b)-(d) The integrated data signal (black solid) compared to theory (blue dashed) with $E_{ext}=50$ mV/cm. Neutrality curve in (b) corresponds to the electron signal in (a). Plasma parameters are (b) $N_{i}=1.1\times10^6, v_{0}=47$m/s, (c,i) $N_{i}=0.23\times10^6, v_{0}=75$m/s, (c,ii) $N_{i}=1.0\times10^6, v_{0}=70$m/s, and (d) $N_{i}=0.92\times10^6, v_{0}=95$m/s. Electron signals are the average of 50 shots with characteristic statistical uncertainty shown in (b).}
\label{esig}
\end{figure}

A typical emission curve from our experiment over the entire lifetime of the plasma is shown in Fig.~\ref{esig}(a), showing a prompt peak of electrons emitted during plasma formation, followed by electrons continuously emitted during the expansion phase. In principle, this signal contains information about the electron distribution and evaporation processes.  At a given time, each electron is bound to the plasma in a potential well of finite depth, created by the combination of the slow-moving ion cloud, the applied electric field and the other electrons.  Since the electron collision times are short compared to the plasma expansion, electrons will evaporate out of the plasma until a quasi-equilibrium is established.  This evaporation differs from that for neutral atoms in that the trapping potential depends on the electron number; for each electron lost, the plasma potential well binding the remaining electrons deepens.  As the density decreases from expansion, the external field can penetrate farther into the plasma, spilling more electrons and forcing the evaporation to continue.

In this paper, we separate the effects of the external electric field on electron emission and find that this consideration alone can reproduce the basic shape of the electron emission curve.  We use a zero temperature approximation, and present an electrostatic calculation that gives the electron spatial distribution for a fixed ion density and applied field.  This gives the maximum number of electrons that can be held in the plasma in the absence of evaporation.  Applying this calculation at all times during expansion allows us to predict the flux of cold electrons from the plasma, which we find is in excellent agreement with the observed electron signal.

The creation of an ultracold plasma in our experiment has been described in previous work \cite{killian1999}.  Briefly, we prepare a sample of $5\times10^6$ metastable Xe atoms in a magneto-optical trap.  The atomic density is roughly Gaussian with an rms radius of $\approx 300\mu m$.  We create a plasma using two-photon photoionization with one photon at the cooling transition (882 nm) and the second from a tunable pulsed dye laser (514 nm).  The initial energy given to the electrons, $E_{e}$, is controlled by tuning the energy of the pulsed laser above the ionization limit.  After creation, the plasma loses a few percent of the electrons until a large enough charge imbalance exists to trap the remaining electrons, which quickly thermalize (the electron thermalization time is $\approx$ 20 ns at the center of the plasma at formation densities) and establish a quasi-equilibrium.  This initial loss is represented by the prompt peak in the signal in Fig.~\ref{esig}(a).  During the expansion phase, the ion distribution maintains a Gaussian density following a self-similar expansion described by $\sigma^2 (t) = \sigma^2 (0) + v_{0}^2 t^2$ with a rate of $v_{0}=50-100$m/s, driven by the thermal electron pressure.  

%
%

\begin{figure}[b]
\vskip  -.20in
\includegraphics[width=3.4in]{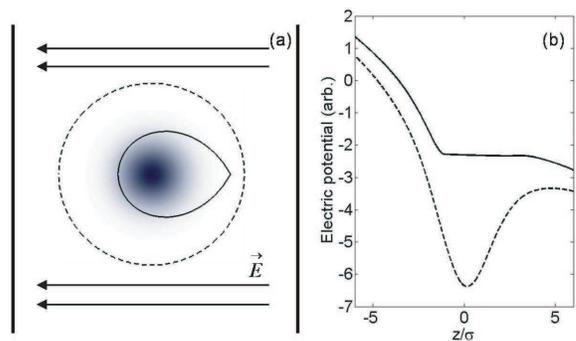}
\vskip -0.2in
\caption{ (a) A schematic of our setup including a contour plot of ion density and electric grids.  The  solid line shows the sharp edge of the cold electron cloud, and the dashed line is the 3$\sigma$ radius of ion cloud. (b) The potential along the symmetry axis of the ion distribution plus external field (dashed) and the total distribution including cold electrons (solid). }
\label{schematic}
\end{figure}

Two wire mesh grids are located 1.5 cm on either side of the plasma.  An electric field (typically just 5-10 mV/cm) applied between these grids directs electrons out of the plasma region, through a third grid and onto a microchannel plate detector.  By controlling the voltages on the grids and detector, we can detect either electron or ion emission from the plasma.  A schematic showing a plasma polarized by this field is shown in Fig.~\ref{schematic}(a).

As the plasma expands, the electron and ion densities decrease and the electron temperature decreases following adiabatic expansion. Previous measurements show the electron temperature scales approximately as $t^{-4/3}$ for times $>15\mu$s \cite{fletcher2007}.  The combined effect means the thermalization times for electrons steadily increase but remain less than a few $\mu$s over the entire plasma lifetime.  This allows us to assume that the electrons maintain a quasi-equilibrium within the continually decreasing potential well.  

For our calculation, we first consider a fixed point in time and allow the electrons to move and establish equilibrium within the stationary ion density.  An example of a Gaussian ion cloud tipped by a uniform external field is shown in Fig.~\ref{schematic}(b) (dashed curve).  We treat the electrons as a zero-temperature fluid, and seek the maximum number of electrons such a potential can hold and the resulting spatial distribution.  The zero-temperature approximation requires a net zero electric force on all electrons in equilibrium, so the maximum number of electrons can be determined by adding electrons until the combined electron and ion potential becomes flat (Fig.~\ref{schematic}(b) (solid curve)).  In the absence of an external field, the solution is an electron density that exactly matches the ion density everywhere in space.  But an external field reduces the number of electrons that can be held, pushing the extracted electrons away from the system.  The electron density will still exactly match the ion density in some region of space with a sharp boundary and be zero outside that boundary.  The shape of the boundary is defined such that the ion density outside the boundary has the correct shape to exactly cancel the external field everywhere within the electron cloud, resulting in a net zero force on all electrons.

\begin{figure}[t]
\includegraphics[width=3.4in]{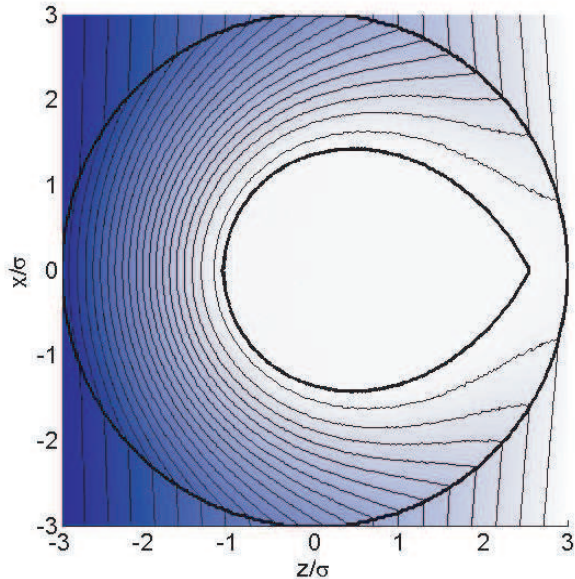}
\vskip -0.1in
\caption{ A potential map of the plasma showing lines of equipotential for $\alpha$ = 0.0035, which corresponds to an electron fraction of 0.47.  The thick solid lines are the boundaries of the ion and electron clouds.  The potential is constant everywhere inside the electron cloud boundary. }
\label{potentialmap}
\end{figure}

To find the proper shape of the electron cloud, we use a numerical algorithm first used by B. Breizmand and A. Arefiev to analyze a similar problem in laser-irradiated cluster physics \cite{arefiev2003, arefiev2005}.  We define a boundary to the electron cloud that is initially at the edge of the ion distribution.  We take the ion density to be a Gaussian with a cutoff at 3 $\sigma$.  The cutoff reflects the predicted existence of an ion density spike near the 3 $\sigma$ point of the expanding plasma that arises from the initial charge imbalance \cite{robicheaux2003, pohl2004}.   We calculate the electric field at the boundary and then displace the boundary by an amount proportional to the local value of the field.  After repeating this process over many iterations, the boundary motion slows down and eventually stops as it reaches the case of zero field at all boundary points.  Inside the boundary the electron density exactly matches that of the Gaussian ions and the integration of this gives the total electron number.  An example of the electron boundary shape is given in Fig.\ref{potentialmap}.

We perform this calculation at all times during expansion, assuming the ion cloud center is fixed and the size follows the ideal self-similar expansion.  As the density decreases, the electric field has a stronger polarizing effect such that fewer and fewer cold electrons can be held in by the ions.  Eventually the density is low enough that the electric field dumps all electrons.  We calculate the maximum number of cold electrons as a function of time (ion density).  

The shape of the boundary and the fraction of electrons to ions, $N_e/N_i$, depends only on the ratio of the external field, $E_{ext}$, to the characteristic field of the ions.  We find $N_e/N_i$ for one set of dimensionless values of 
	\[\alpha = \frac{E_{ext} \epsilon_0 \sigma^2}{e N_{i}},
\]
where $\epsilon_0$ is the electric constant, $e$ is the electron charge, and $N_i$ is the number of ions.  Measurements of the plasma expansion velocity and density are sufficient to compare to experiment. 

Figure~\ref{esig}(b) shows the fraction of electrons in the plasma as a function of time, found by integrating the signal in Fig.~\ref{esig}(a), and the corresponding calculated cold electron fraction.  The calculation predicts the general shape and timescale of the signal over a wide range of initial parameters.  As expected, the calculation overestimates the plasma neutrality during the prompt loss of thermal electrons, the discrepancy greater for higher $E_{e}$ and lower $N_{i}$ \cite{killian1999}.  The expansion velocity of the plasma is measured using the plasma response to an rf field \cite{kulin2000}, and the number of plasma ions is left as a fit parameter.  Fitted $N_{i}$ values fall within the uncertainty of previous independent measurements made using MOT diagnostics and other plasma processes.  

\begin{figure}[t]
\includegraphics[width=3.4in]{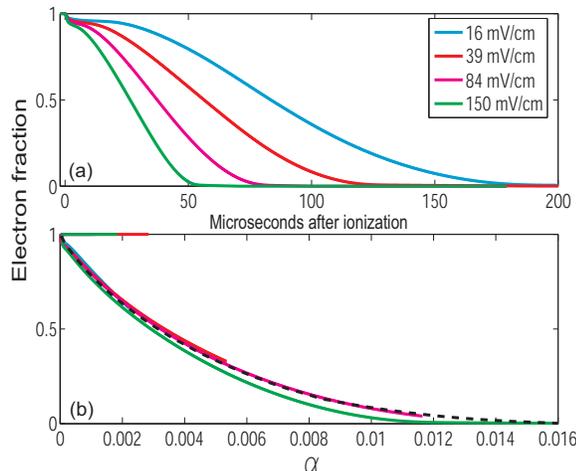}
\vskip -0.2in
\caption{(a) Integrated electron signals for varied external field, and (b) the same signals plotted vs the normalized $\alpha$. The black dashed line is the theoretical result. Data signals are the average of 50 shots with $E_{e}/k_{b}=100$K and $N_{i}=1\times10^6$. The ion cloud starts crossing the grids at 71$\mu$s.}
\label{efield}
\end{figure}

Figure~\ref{efield} shows a direct comparison of data taken with different strengths of electric field.  Stronger fields shorten the plasma lifetime, but plotting the signals using our normalized coordinate $\alpha$ shows a similar rate of electron loss for all.  Electron signals taken for varied initial density and initial electron energy are also found to follow this universal curve.  For initial energies below $E_{e}/k_{b}\sim60$K the shape of the observed signals does not match as well to the theory.  At these low temperatures, an increased three-body recombination rate converts as much as 15\% of the plasma ions into Rydberg atoms, \cite{pohl2004,robicheaux2003} (our calculation has assumed a constant ion number) and the model of self-similar Gaussian expansion is less accurate \cite{gupta2007}.  

Late in its lifetime, the plasma is large enough for the edges of the ion cloud to start crossing the mesh grids.  By monitoring the ion current signal, we determine that the first ions start crossing the grid when the grid position is between 3 $\sigma(t)$ and 4 $\sigma(t)$.  The ion signal also shows a sharp turn-on, indicating a sharp edge to the ion density, which supports our assumption of a Gaussian density with a sharp drop at $\approx$3 $\sigma$.  For times after this, a more correct density includes a planar cut-off at the locations of the grids.  In Fig.~\ref{efield}(b), we compare only the times before the ions start crossing the grid.  Practically this makes little difference, as the grids only cut off the low density outer regions of the plasma.  We have performed the cold electron calculation with the grids included for a large and slow ($9\times10^5$ ions and $v_{0}=60$ m/s) plasma with a low electric field (15 mV/cm), where the lifetime is long.  The result gives lower neutrality for all times after the plasma starts crossing the grid, but the differences are less than 5 percent compared to the case where we ignore the grids.

Above 200 mV/cm, the signal scaled with $\alpha$ begins to diverge from the universal curve seen in Fig.~\ref{efield}(b).  For strong fields, the neutrality drops quickly enough that it alters the ion expansion, both from the linear acceleration of free ions in the applied field and from the Coulomb repulsion between the ions.  The Gaussian density approximation then becomes invalid.  Residual electric fields in our vacuum chamber prevent a good fit down to arbitrarily small field, and ultimately limit the maximum observable plasma lifetime.  Stray fields along the grid axis can be canceled with slight adjustments to the grid voltages.  We monitor the total number of electrons detected over the full lifetime and see a drop from all electrons detected to none as one grid voltage is changed over a 25 mV range.  The midpoint of this transition should be close to zero axial field and typically comes with a 20-30 mV difference between the grids.  This difference is taken into account in our reported field values.  Stray transverse fields are not canceled, but we estimate them to be no greater than 20 mV/cm based on the quality of our fits near zero axial field.

Expanding ultracold neutral plasmas are systems in quasi-equilibrium, sharing this aspect with a wide variety of physical systems from atomic to galactic clusters.  We have presented data of electrons evaporating from the expanding plasma that is in excellent agreement with a zero-temperature electron model, where the electrons arrange themselves in an asymmetric density distribution to exactly cancel the ion plus external electric fields. This result does not directly address the question of the form of the electron energy distribution function, but rather serves as a foundation that can be used in further experiments.  Experiments using the electron leakage to probe electron energies must find a way to separate or eliminate the spilling of electrons from thermal evaporation.  

Eliminating the effect of the external field might be done by dumping a significant percentage of the electrons at early time to create a deep potential well, and watching for electrons that escape in the following time interval when the cold electron flux should be zero.  Alternatively, we can imagine decreasing the electric field continuously to counter the effect of expansion and prevent any cold electron loss (i.e. keep $\alpha$ constant) and detect the emitted thermal electrons.  Our ability to directly detect the emitted particles as well as perturb the potential in a time dependent manner suggests that UNPs are a good candidate system to address fundamental questions about systems in quasi-equilibrium.

\begin{acknowledgments}
We are grateful to A. Arefiev for his help with the electron boundary calculation.  This work was partially supported by the National Science Foundation PHY-0714381.
\end{acknowledgments}

\bibliography{esigaip}

\end{document}